\documentclass[twocolumn,showpacs,preprintnumbers,amsmath,amssymb]{revtex4}

\usepackage{graphicx}% Include figure files
\begin{document}

% THIS IS THE TITLE OF THE ARTICLE:
\title{Some Statistical Mechanical Properties of Photon Black Holes}

% \shorttitle{Photon black holes}

\author{Hernandez, X.$^{1}$}
\email[Email address: ]{xavier@astroscu.unam.mx}
\author{Lopez-Monsalvo, C. S.$^{1}$}
\email[Email address: ]{cslopez@astroscu.unam.mx}
\author{Mendoza, S.$^{1}$}
\email[Email address: ]{sergio@astroscu.unam.mx}
\author{Sussman, R. A.$^{2}$}
\email[Email address: ]{sussman@nuclecu.unam.mx}
\affiliation{$^1$Instituto de Astronom\'{\i}a, Universidad Nacional
                 Aut\'onoma de M\'exico, AP 70-264, Distrito Federal 04510,
	         M\'exico \\
	     $^2$Instituto de Ciencias Nucleares, Universidad Nacional
	         Aut\'onoma de M\'exico, AP 70-543, Distrito Federal 04510,
		 M\'exico }

\date{\today}

\begin{abstract}
We show that if the total internal energy of a black hole is constructed
as the sum of $N$ photons all having a fixed wavelength chosen to
scale with the Schwarzschild radius as $\lambda=\alpha R_{s}$, then
$N$ will scale with $R_{s}^{2}$.  A statistical mechanical calculation
of the configuration proposed yields \( \alpha = 4 \pi^2 / \ln(2) \)
and a total entropy of the system $S=k_{B}N \ln(2)$, in agreement with
the Bekenstein entropy of a black hole .  It is shown that the critical
temperature for Bose-Einstein condensation for relativistic particles of
$\lambda=\alpha R_{s}$ is always well below the Hawking temperature of a
black hole, in support of the proposed internal configuration.  We then
examine our results from the point of view of recent loop quantum gravity
ideas and find that a natural consistency of both approaches appears.  We
show that the Jeans criterion for gravitational instability can be
generalised to the special and general relativistic regimes and holds for
any type of mass--energy distribution.
\end{abstract}

% PACS numbers:

\maketitle

\section{Introduction}

In general, black holes are defined uniquely by their mass, angular
momentum and charge (cf. \cite{chandra}). In this paper we shall
deal exclusively with Schwarzschild black holes, where the angular
momentum and charge are both zero. These black holes of mass $M$ are
understood as systems defined uniquely by the condition $R \leq R_{s}$
and have the following properties:

\begin{equation}
  R_{s}= \frac{ {2 G M} }{ {c^{2}} },
\end{equation}

\begin{equation}
  \frac{ S }{ k_{B} } = \frac{ A_{BH} }{ 4 A_{P} },
\end{equation}

\begin{equation}
  T_{BH}= \frac{ hc }{ 8 \pi^{2} k_{B} R_{s} }.
\end{equation}

\noindent Equation (1) defines the Schwarzschild radius in terms of the
mass. Relation (2) establishes the entropy \( S \) of the system as $1/4$
of the horizon area $ A_{BH} = 4 \pi R_{s}^{2}$, in units of the Planck
area  $ A_{P} = \hbar G /c^{3}$. Equation (3) states that the black
hole radiates as a black body of the given temperature \( T_{BH} \).
This emission process implies a loss of energy for the system, which
results in an evaporation rate for the black hole given by: $ dM/dt
\propto (\hbar c^{4}/G^{2}) M^{-2} $.

Notice that $dS/dM = (c^{2}T)^{-1}$, fixing the internal energy of the
black hole as $U=Mc^{2}$. Equation (1) has been about in speculative form
since the 18th century, and was given a firm theoretical footing within
the framework of general relativity during 1930s. Equations~(2)-(3) are
the result of the vigorous development in black hole thermodynamics of
the 1960s and 1970s by various authors, notably Bekenstein and Hawking (cf.
\cite{nov} and references therein).

It has been suggested (e.g. \cite{'t Hooft}, \cite{Suss} or \cite{Bousso}
for a review) that for any physical system equation (2) should hold
always, with $\leq$ replacing the equality, which in turn should hold only in
the black hole regime. This is termed the holographic principle, from
the fact that the information content of an object would be limited
not by its 3D volume, but by its 2D bounding surface. The interesting
connection implied between quantum mechanics through $A_{P}$ and gravity
through the particle horizon, has raised the hope that the validity of
the holographic principle would yield important clues regarding quantum
theories of gravity. In this sense, even a heuristic study as to the
possible origin of this principle should prove valuable.

We study the behaviour of a classical black body photon gas as it is
compressed into a black hole, and propose a simple model for such
a system using only photons confined to the Schwarzschild radius
at their lowest possible momentum level. Two parameters determine
the model, with a restriction only on the product of both of them.
A formal statistical mechanical calculation is given through which
these parameters are determined, leading to agreement with all black
hole structural properties of equations (1-3).
The study of a self--gravitating photon sphere (a geon) was first
introduced by  Wheeler \citep{geons} in 1955.  Much development in this area
has grown since then related to different physical properties of boson
stars (see e.g. \citep{jetzer,schunck}) and their stability. 

Given recent proposals of a physical model for a black hole interior
within the framework of loop quantum gravity \cite{Rov}, we analyse our
model in this context. Taking the view that the Bekenstein entropy has an
statistical mechanical origin in terms of counting states on the surface
defined by the Schwarzschild radius of a black hole, canonical quantum
gravity has yielded scenarios in which this entropy can be derived from
first principles.  We find no inconsistencies with the quantum gravity
approach, which in fact allows us to explicitly and independently
re-evaluate the parameters introduced in the quantum simple model,
obtaining the same results.

\section{Classical Limit}

Most of the material in this section can be found elsewhere, e.g. \cite {Sorkin}, 
it is reproduced here for context. 
For a photon gas having a black body spectrum the following well known
relations define the total electromagnetic energy $E_{EM}$ and entropy $S$
in terms of the volume and temperature:

\begin{equation}
  E_{EM}={{\pi^{2}}\over {15}} {{(k_{B}T)^{4}}\over {(\hbar c)^{3}}}
    {{4 \pi R^{3}}\over {3}},
\end{equation}

\begin{equation}
  {{S}\over{k_{B}}} ={{4 \pi^{2}} \over {45}} {{(k_{B}T)^{3}} \over {(\hbar
    c)^{3}}} {{4 \pi R^{3}} \over {3}},
\end{equation}

\noindent for a spherical region of radius $R$.

If we think of an ideal adiabatic wall enclosing this spherical region,
we immediately obtain the well known scaling of $T \propto R^{-1}$,
and we can eliminate $(k_{B}T)$ from equation(4) in favour of $S$ and \(
R \) to obtain

\begin{equation}
E=C {{\hbar c } \over {R}} \left( {{S} \over {k_{B}}}\right)^{4/3},
\end{equation}

\noindent where $C$ is a numerical constant of order unity. If we think
of the contraction as proceeding into the black hole regime, we would
think of the radius as reaching $R_{s}$, which in this case would yield

\begin{equation}
  R_{s}={{2 G E_{EM}} \over {c^{4}}}.
\end{equation}

  In equation (7) we have used equation (1), replacing $M$ for
$E_{EM}/c^{2}$. Substitution of $E_{EM}$ from equation
(6) into equation (7) leads to

\begin {equation}
  \left( {{S} \over {k_{B}}}   \right)^{4/3} = C^{\prime} {{A_{BH}}
    \over {A_{p}}},
\end{equation}

\noindent where $C^{\prime}$ is a numerical factor of order unity.
Two interesting conclusions are immediately evident from this last
equation. Firstly, it is obvious that for any black body photon gas having
$R_{s} > R_{p}$, a Schwarzschild radius larger than the Plank length,  
the holographic principle will be valid throughout
the contraction process, as the horizon area will always be larger than
$A_{BH}$. Second, that the classical equations for the diluted photon gas,
equations (4-5), cannot be valid into the black hole regime, since the
required relationship for the entropy of such an object is equation (2)
and the exponent of equation (8) is $4/3$ and not $1$. The inconsistency
of equation (8) with equation (2) signals that physical processes which
the system being modelled surely experiences, such as pair creation at
high temperatures and quantum effects related to the dimensions of the
system being comparable to the typical de Broglie wavelengths of the
photons, are not been taken into account by the classical description
of the photon gas through equations (4-5). 
So far, we have assumed that the photon gas was being compressed by some
external agency, however, if it is to form a self gravitating object, an
equilibrium configuration should exist, and possibly a collapse beyond
this. This point can be estimated by evaluating the Jeans length $R_{J}$
of the problem for a sound speed $v_{s}=c/\sqrt{3}$, 

\begin{equation}
  R_{J}={{c}\over {(3 G \rho)^{1/2}}  }.
\label{hola}
\end{equation}

  In this context, the mass density \( \rho \) is equivalent to $E_{EM}
/ V $.  Notice that equation~\eqref{hola} is derived directly from
Einstein's equations, where the pressure term is directly $\mathrm{d }
E/ \mathrm{d} V$, hence no assumption of particle interactions is being
made (see the appendix for details on this).  From equation~\eqref{hola}
we see that since $\rho$ scales with $T^4$, $R_{J}$ will scale with
$T^{-2}$, which is interesting given that under adiabatic conditions
the radius of the system will scale with $T^{-1}$. This means that
gravitational instability will occur, i.e. $R>R_{J}$ above a certain
critical temperature, below a certain critical equilibrium radius
$R_{c}$. The situation becomes increasingly unstable in going towards
smaller radii and larger temperatures. In general, for a fluid of
mass--energy $M$, radius $R$ and sound speed $v_{s}$,

\begin{equation}
  R_{J}=v_{s} \left( { {4 \pi R^{3}} \over {3 G M} } \right) ^{1/2},
\end{equation}

\noindent which expressing M in terms of the Schwarzschild 
radius through equation (1) reads,

$$
  R_{J}=\left( {{8 \pi} \over {3}}   \right)^{1/2} {{R v_{s}}\over {c}}
    \left( {{R} \over {R_{s}}} \right)^{1/2}.
$$

\noindent If we take the critical condition $R=R_{J}$, the previous
relation gives

\begin{equation}
  {{R_{J}} \over {R_{s}}} = \left( {{3}\over{8 \pi}}\right) \left( {{c}
    \over {v_{s}}}\right)^{2}.
\end{equation}

Equation(11) shows that, since for all non--relativistic systems
$v_{s} \ll c$, we should expect $R_{J} \gg R_{s}$. Indeed, for most
astrophysical applications the Jeans radius of a system is many orders of
magnitude larger than the Schwarzschild radius.  However, in going to a
relativistic fluid, the condition $v_{s} \simeq c$ will apply, leading to
$R_{J} \simeq R_{s}$. In other words, the self-gravitating regime will
appear only close to the black hole regime.  For the adiabatic photon
gas we have studied, taking $v_{s}=c/\sqrt{3}$ and equations (4-5) and (9), in
correspondence with equation (11) we obtain also $R_{J} \simeq R_{s}$.
This last result shows that the self-gravitating regime for the photon
gas we have studied does not appear until one is very close to the
black hole regime, at scales where the analysis leading to equation (8)
already showed that the structure equations for the gas (4) and (5) are no
longer valid. In any case, the analysis following equation (9) together
with  equation (11), strongly suggests that any self-gravitating photon
gas will be very close to catastrophic collapse and black hole formation.

The above results are in fact valid into the regime where the self-gravity of the
radiation field is important, as shown by \cite{Sorkin}, instability sets in for
$R<2R_{s}$, but equilibrium maximum entropy configuration exist above this radius, 
which however also show the scaling of equation (8), c.f. their results following 
their equation (41).

\section{Quantum limit}

 The gravitational collapse and transition between the classical
regime of section~II and a black hole will not be treated explicitly.
Advances in that direction can be found e.g. in \cite{Sorkin} \cite{Pavon}
and \cite{Ding}.

  Being subject to the extreme gravitational regime of $R \rightarrow
R_{s}$, it is reasonable to expect that the photons will be highly limited
in momentum space. At this point we introduce as a hypothesis that all
photons will have a wavelength $\lambda=\alpha R_{s}$, with $\alpha$ a
numerical constant.  We can evaluate the internal energy of the system as:

\begin{equation}
  E_{EM}={{N h c } \over {\alpha R_{s}}},
\end{equation}

\noindent where $N$ is the total number of photons. Establishing a
correspondence between $E_{EM}$ and the internal energy of a black hole as
$E_{EM}=Mc^{2}$,
and using equation (1) to express $M$ in terms of $R_{s}$, the above
expression yields:

\begin{equation}
N={{\alpha} \over {16 \pi^{2}}} {{A_{BH}} \over {A_{p}}} .
\label{eq-for-n}
\end{equation}

  If we think of a correspondence between the total entropy of the system
and the total photon number given by $S/k_{B}=\beta N$, with $\beta$
a proportionality constant expected to be of order unity, 
we find by comparison to equation (2) that the
configuration we propose will satisfy all required black hole properties
if the condition $\alpha \beta = 4 \pi^{2}$ is satisfied.

  We can compute $ \beta$ directly by calculating the entropy of the
proposed system from first principles, through the thermodynamic potential
$\Omega$ given by:

\begin{equation}
\Omega=k_{B} T \sum_{k} \ln \left( 1 - e^{[\mu -\epsilon_{k}] /k_{B}T} \right),
\end{equation}

\noindent where the summation is over quantum states, $\mu$ is the
chemical potential, and $\epsilon_{k}$ is the energy of the $k-th$
state. Since the total number of components of the system is given by:

\begin{equation}
N=\sum_{k} \left({{1} \over {e^{[\epsilon_{k} - \mu] /k_{B}T}-1}  }\right),
\end{equation}

\noindent and given that in the system proposed all photons have the
same energy, we can write

\begin{equation}
N={{N} \over {e^{[\epsilon - \mu] /k_{B}T}} -1}.
\end{equation}

\noindent Note that each photon is assumed to be in a distinct
detailed quantum level, and hence the analogy with a condensate is not
complete. Now,

\begin{equation}
[\epsilon - \mu] / k_{B} T =\ln(2),
\end{equation}

\noindent which when substituted back into equation~(14) gives:

\begin{equation}
\Omega=-k_{B} T N \ln(2).
\end{equation}

  This last result now yields the entropy for the system through $S=
-\partial \Omega / \partial (k_{B}T) |_{V,N}$ as $S=N \ln(2)$, providing a
justification for the assumption of $S/k_{B}=\beta N$ made above, in terms
of simple statistical physics arguments.  We hence obtain $\beta=\ln(2)$
and $\alpha =4 \pi^{2} /\ln(2)$.  Note that the previous results are
of general validity for bosons, the case of photons is obtained with \(
\mu = 0 \).

  We also note that given the restriction of a fixed total internal
energy, in the hypothesis that this is to be the sum of the energies
of $N$ photons, the maximum entropy state will be the one with the most
photons. In that case, all photons are at their lowest possible energy
$\lambda=\alpha R_{s}$.  Hence, the configuration proposed is also a
maximum entropy state and is suggestive of a micro--physical origin for
black hole entropy.

  It has been argued \citep{bekenstein82} that the shortest scale that can
enter into any physical theory is the Planck length.  Although so far we
have been working under the assumption of macroscopic black holes, we can
extrapolate to the very small scales as follows.  In the context of the
ideas presented here a natural lower limit for the Schwarzschild radius
of order the Plank length appears by setting \( N=1 \) in equation~(13),
a single photon black hole.  The energy \( e \) associated to this single
photon  is  \( e \approx \hbar c / ( G h / c^3 )^{1/2} \approx 10^{28}
\, \text{eV} \).  

  Note that using equation (1) to substitute \( M \) for \( R_{s} \)
in equation~\eqref{eq-for-n} gives the following quantisation for the
mass \( M_N \) of a black hole in units of the Planck mass:

\begin{equation}
  \frac{ M_N }{ m_p } = \left( \frac{ \pi }{\alpha } \right)^{1/2}
    N^{1/2},
\label{qt01}
\end{equation}

\noindent where \( m_P = \left( \hbar c / G \right)^{1/2} \), a
quantisation suggested already by Bekenstein's entropy equation (2). For
sufficiently small black hole masses,  this equation suggests a discrete
spectrum associated to the transitions \( N = 1 \rightarrow 0 \),
\( N = 2 \rightarrow 1 \), etc.  The change in mass \( \Delta M_N \)
corresponding to a \( \Delta N = 1 \) transition  is given by

\begin{equation}
  \Delta M_N  = M_N  \left\{ \left[ 1 + \frac{ \pi }{\alpha } \left(
    \frac{ m_P }{ M_N } \right)^2 \right]^{1/2} - 1 \right\}.
\label{qt02}
\end{equation}

  In the limit of a macroscopic black hole, where \( M_N \gg m_p \),
the above equation implies that

\begin{equation}
  \frac{ \Delta M_N }{ m_P } = \frac{ \pi }{ 2 \alpha } \, \frac{ m_P }{ M_N
    }.
\label{qt03}
\end{equation}

\noindent It is interesting to note that \( c^2 \Delta M_N \) approximately
corresponds to \( k_B T_{BH} \) for a black hole mass \( M_N \).

  If we now identify the time scale \( \Delta t \) for the mass loss \(
\Delta M_N \) to take place with the limit of the Heisenberg 
uncertainty principle, we can set up \( \Delta t \sim \hbar  
/ c^2 \Delta M_N \).  Under the above considerations, the mass evaporation rate
for a black hole is

\begin{equation}
  \frac{ \Delta M_N }{ \Delta t } \propto \frac{ \hbar c^4 }{ G^2 } \,
    \frac{ 1 }{ M_N^2 },
\label{qt04}
\end{equation}

\noindent which is within a numerical constant of the standard evaporation
rate for a black hole \citep{nov}, seen here as the macroscopic limit of an
intrinsically quantum process.

  If the structure of the photon configuration we are describing is in any
way related to quantum phenomena akin to Bose-Einstein condensation, we
should expect the temperature to lie well below the critical temperature
$T_c$ for Bose-Einstein condensation. This, for relativistic particles
can be calculated in an analogous way to well known Bose-Einstein
critical temperatures for non-relativistic particles (e.g. \citep{landau}), 
by integrating the expression

\begin{equation}
dN={{g V p^{2} dp} \over {2 \pi^{2} \hbar^{3}  \left[ e^{(\epsilon -
  \mu)/k_{B} T}  \right]  }},
\end{equation}

\noindent for $\mu=0$, $g=2$ (photons) and in this case $\epsilon=pc$, giving

$$
N={ {V (T_{c} k_{B})^{3} } \over {\pi^{2}\hbar^{3} c^{3} } }
\int_{0}^{\infty} { {z^{2} dz} \over {e^{z}-1}},
$$

\noindent where $z=\epsilon/(k_{B}T)$. Evaluation of the above integral
yields $2.202$ and so, using $V=(4 \pi/3) R_{s}^3$ leads to

\begin{equation}
N=\left( \frac{8.808}{3 \pi} \right) {{R_{s}^{3} (T_{c} k_{B})^{3} } \over { (\hbar c)^{3}}},
\end{equation}

  If we now use the expression for $N$ in equation~\eqref{eq-for-n},
and writing $T_{c}$ in units of $T_{BH}$ for a black hole of radius 
equal to the Planck length \( R_p \), which correspond to a
temperature $T_{BHp}$, we get

\begin{equation}
\left( { T_{c} }\over {T_{BHp}} \right)^{3}= \left({6 \pi^{3}}\over
{1.101}\right) {{\alpha R_{p}}\over {R_{s}}},
\end{equation}

\noindent which for $\alpha = 4 \pi^{2}/\ln(2)$, as determined through the statistical mechanical
calculation shown above  yields,

\begin{equation}
\left({T_{c}} \over {T_{BHp}} \right) =21.3 \left( {R_{p}} \over
  {R_{s}}\right)^{1/3}.
\end{equation}

\begin{figure}
\begin{center}
\includegraphics[height=8.5cm]{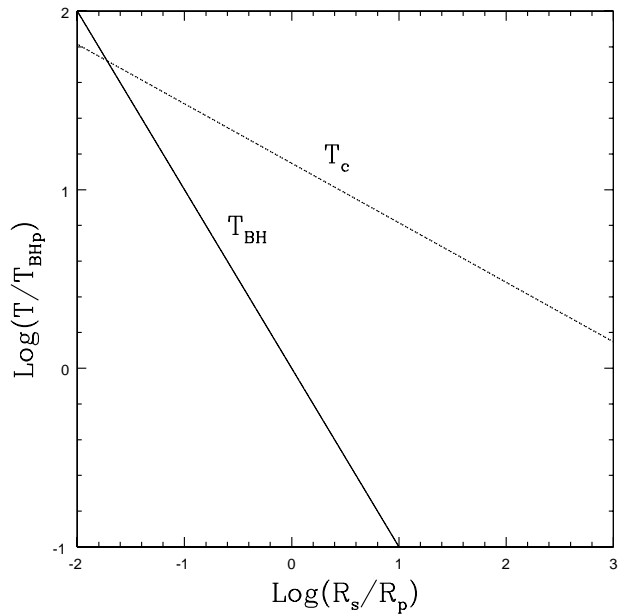}
\end{center}
 \caption{The solid line shows the black hole temperature, in units of
this quantity for a black hole having a Schwarzschild radius equal to
the plank length, as a function of the Schwarzschild radius, in units
of the Plank length.  The dotted line gives the critical temperature for
Bose-Einstein condensation of photons in a black hole, equation (26),
in the same units as the solid curve, as a function of the same quantity.}
\end{figure}

As $T_{c}$ scales with $R_{s}^{-1/3}$ and $T_{BH}$ scales with $R_{s}^{-1}$,
it is clear that for all black holes larger than a certain
limit, the condition $T_{BH} \ll T_{c}$ will be satisfied. A comparison
of both temperatures is shown in figure 1, as a
function of $R_{s}$, from which we see that $T_{c}$ is already
over an order of magnitude larger than  $T_{BH}$ at  $R_{s}= R_{p}$. Any
realistic black hole will be at a temperature much lower than the critical
temperature for Bose-Einstein condensation for photons, showing the
internal consistency of the model.

The physics described so far is clearly highly idealised, however in a
core collapse process within a massive star, as for the central region $R
\rightarrow R_{s}$, the typical speeds and $v_{s}$ of the constituent
particles must necessarily tend to $c$, with de Broglie wavelengths
not larger than the Schwarzschild radius.

At this point, quantum effects similar to Bose-Einstein condensation
could take place, packing all (or most) photons into the lowest
energy state. In this sense, typical wavelengths of the order of
the Schwarzschild radius would be expected, as longer wavelengths
would be prohibited by the containment imposed by gravity, and shorter
wavelengths would imply an expansion in momentum space. In this sense, the
identification we have maintained of the constituent particles as photons
is shown to be largely arbitrary, any relativistic bosons will yield
essentially identical conclusions.

\section{Loop Quantum Gravity Approach}
\label{loop}

 We now explore the ideas of the previous sections within the framework of
loop gravity.  In particular, we will see that this allows us an independent
re-evaluation the constants $\alpha$ and $\beta$ established in the last section, with
the aid of some general results from loop quantum gravity.

  The loop quantisation of $3+1$ general relativity is described in
terms of a set of spin network states which span the Hilbert space on
which the theory is based. These spin network states are labelled by
closed abstract graphs with spins assigned to each link and intertwining
operators assigned to each vertex.

  A resent result that follows from the theory is that if a surface
$\Sigma$ is intersected by a link $\ell_i$ of a spin network carrying
the label $j_i$ it acquires an area \cite{ashtek1,rov1}

\begin{equation}
\label{area1}
A_{\Sigma}(j_i)=8\pi A_p \gamma \sqrt{j_i(j_i+1)},
\end{equation}

\noindent where $\gamma$ is the Immirzi parameter.

  Let us now consider for our purposes that the horizon $\Sigma$ is
intersected by a large number $N_\ell$ of links. Each intersection
with $\Sigma$ represents a  puncture. In the limit of large $N_\ell$,
one can say that each puncture is equipped with an internal space $H_j$
(the space of all flat $U(1)$ connections on the punctured sphere) of
dimension \cite{alexan1}

\begin{equation}
\label{dim1}
\dim H_j=2j+1.
\end{equation}

Each puncture of an edge  with spin $j$ increases the dimension of the
boundary Hilbert space by a factor of $2j+1$.  Under these considerations,
it follows that the  entropy can be calculated by

\begin{equation}
\label{entropy}
S({j_p})=\ln\left ( \prod_{p} \dim H_{j_p} \right ).
\end{equation}

Statistically, the most important contribution comes from those
configurations in which the lowest possible spin $j_{min}$ dominates,
so we can write the entropy (\ref{entropy}) as

\begin{equation}
\label{entropy2}
S(j_{min})=N_\ell \ln (2j_{min}+1),
\end{equation}
where $N_{\ell}$ is given  by

\begin{equation}
\label{Npunc1}
N_{\ell}=\frac{A_{BH}}{A_\Sigma(j_{min})}.
\end{equation}

  Due to the fact that the assumed gauge group of loop quantum gravity is
SU\( (2) \), then it follows that \( j_\text{min} = 1/2 \), and so the 
Immirzi parameter is given by \cite{rov1}

\begin{equation}
  \gamma=\frac{\ln 2}{ \pi \sqrt{3} },
\label{immirzi}
\end{equation}

\noindent and (\ref{Npunc1}) becomes

\begin{equation}
\label{Npunc2}
N_\ell=\frac{1}{4 \ln 2}\frac{A_{BH}}{A_p}.
\end{equation}

  The number of links \( N_\ell \) associated to the particles we are
dealing with in this article  must be proportional to the number of
particles \( N \). The simplest possible configuration is the one in
which the proportionality factor is equal to unity and so,

\begin{equation}
\label{ansatz}
  N_\ell=N.
\end{equation}

\noindent Using this relation and equations~\eqref{eq-for-n} and
(\ref{Npunc2}) we can evaluate $\alpha$ to obtain

\begin{equation}
\label{alfa}
\alpha=\frac{ 4 \pi^2 }{ \ln 2 }.
\end{equation}

  The parameter $\beta$ previously defined through the relation $S=\beta
N$ can now be re-derived independently through equation (\ref{entropy2})
to yield:

\begin{equation}
\label{beta}
\beta=\frac{N_{\ell}}{N}\ln(2j+1)_{j=1/2}= \ln 2.
\end{equation}

From (\ref{alfa}) and (\ref{beta}) we can see that the product $\alpha
\beta = 4\pi^2$ as required by the considerations on section III.  It is
interesting that the model proposed in the previous sections is seen
not to be in conflict with a loop gravity approach, which indeed allows
to re--evaluate the scaling parameters introduced earlier independently
and in accordance with the expectations of the physics discussed above.

\section{Conclusions}

A photon gas contained within an adiabatic enclosure will satisfy the
holographic principle, at least until before reaching the black hole
regime, which approximately coincides with the self-gravitating condition
and where the classical description is no longer valid.

A configuration where all photons have the same energy within
$R=R_{S}$ can be constructed to satisfy all black hole conditions.

This configuration gives rise to a discrete evaporation spectrum for \(
R_S \) close to the Planck length, and in the macroscopic limit allows a
re-derivation of the standard black hole evaporation rate.

Our results are consistent with the loop quantum gravity scheme,
satisfying the constraints required to be in agreement with
the Bekenstein-Hawking entropy for a black hole.

A comparison of both regimes studied is highly suggestive of a heuristic
proof of the holographic principle, as any real system would require an
increase of entropy (at fixed energy and volume) to be turned into the
photon gas we have studied.

\section{Acknowledgements}
X. Hernandez acknowledges the support of grant UNAM DGAPA (IN117803-3),
CONACyT (42809/A-1) and CONACyT (42748).  C. Lopez-Monsalvo thanks
economic support from UNAM DGAPA (IN119203).  S. Mendoza gratefully
acknowledges financial support from CONACyT (41443) and UNAM DGAPA
(IN119203).  We thank the anonymous referee for his comments and 
suggestions which improved the final version of the paper.

\appendix*

\section{Relativistic Jeans criterium for gravitational instability}

  In order to show that the Jeans gravitational instability limit is
valid in the general relativistic regime, let us proceed as follows.
The condition of hydrostatic equilibrium in general relativistic fluid
mechanics is obtained by using the fact that the field is static.
This means that one can describe the problem in a frame of reference in
which the fluid is at rest, with all hydrodynamical quantities independent
of time. This also implies that the mixed space and time components of
the metric tensor are null.  Under these assumptions, the equation of
hydrostatic equilibrium is then given by \citep{daufm}

\begin{equation}
  \frac{1}{w} \frac{ \partial p }{ \partial r } = - \frac{ 1 }{ 2 } \frac{
    \partial }{ \partial r } \log g_{00},
\label{eqA1}
\end{equation}

\noindent where \( g_{00} \) is the time component of the metric tensor, \( w
= e + p \) is the enthalpy per unit volume, \( e \) internal
energy density and \( p \) the pressure.  Oppenheimer \& Volkoff
\citep{oppenheimer} showed that equation~\eqref{eqA1} can be brought to 
the form \citep[see e.g.][]{MTW}

\begin{equation}
  \frac{ \mathrm{d} p }{ \mathrm{d} r } = - \frac{ \left( e + p \right) }{
    r \left( r - \frac{ 2 G M(r) }{ c^2 } \right) } \left\{ \frac{ G M(r)
    }{ c^2 } + \frac{ 4 \pi }{c^2} G r^3 \rho \right\},
\label{eqA2}
\end{equation}

\noindent where the mass--energy \( M(r) \) within a radius \( r \)
is given by

\begin{equation}
  M(r) = 4 \pi \int_0^r{ \rho \, r^2 \, \mathrm{d} r },
\label{eqA2b}
\end{equation}

\noindent and \( \rho(r) := e / c^2  \) is the mass--energy density of 
the fluid.  Note that for the case of relativistic and non--relativistic
dust particles, \( M(r) \) represents the mass of particles within radius
\( r \).  For the case of a photon gas, \( M(r) \) is the mass
corresponding to the internal energy. 

  We now assume that the plasma obeys a Bondi--Wheeler equation of state

\begin{equation}
  p = \left( \kappa - 1 \right) e, 
\label{eqA3}
\end{equation}

\noindent with constant index \( \kappa \).  This means that the sound
velocity \( v_\text{s} \) is given by the relation \( v_\text{s}^2
= c^2 \left( \kappa - 1 \right) \) and so, the left hand side of
equation~\eqref{eqA2} can be written as \( \left( v_\text{s}^2 / c^2
\right) \, \mathrm{d} e / \mathrm{d} r \).  Seen in this way, the left
hand side of equation~\eqref{eqA2} no longer represents gradients of
pressure which are in balance with self--gravitational forces related to
the plasma.  Indeed, the balance with the gravitational forces produced
by the plasma is now related to the gradients of its  proper internal
energy density  \( e \) by

\begin{equation}
  \left( \kappa - 1 \right) \, \frac{ \mathrm{d} e }{ \mathrm{d} r } = -
    \frac{ k e  }{ r \left( r - \frac{ 2 G M(r) }{ c^2 } \right) } \left\{
    \frac{ G M(r) }{ c^2 } + \frac{ 4 \pi }{c^2} G r^3 \rho \right\}.
\label{eqA3b}
\end{equation}

  Let us now take the absolute value on both sides of
equation~\eqref{eqA3b}.  To order of magnitude \( \mathrm{d} e /
\mathrm{d}r  \approx e / r \) and \( M(r) \approx \left( 4 / 3 \right)
\pi \, r^3 \, \rho \).  A gravitational instability occurs when the
absolute value of the left hand side of equation~\eqref{eqA2} (or
equivalently equation~\eqref{eqA3}) is less than the absolute value of
its right hand side.  Using all the above statements it follows that
this instability occurs when the radial coordinate \( r \) is such that

\begin{equation}
  r \gtrsim \sqrt{ \frac{ 3 }{ 8 \pi \left( 3 \kappa - 1 \right) } } \
    \frac{ v_\text{s} }{ \sqrt{ G \rho } } := \Lambda_\text{J}.
\label{eqA4}
\end{equation}

  The quantity \( \Lambda_\text{J} \) on the right hand side of
equation~\eqref{eqA4} is of the same order of magnitude as the standard
Jeans length used in non--relativistic fluid dynamics.  In other words, the
criterium~\eqref{eqA4} means that the Jeans criteria for gravitational
collapse is also valid in the relativistic regime as well.

  For the particular case studied in this article, the constant \( \kappa
= 4/3 \) for a photon gas and the Jeans criteria can be applied to it,
giving \( r \gtrsim \left( 1 / 2 \sqrt{ 2 \pi }\right) v_\text{s} /
\sqrt{ G \rho } \, \).

  We see than the Jeans criterion can be generalised beyond an equilibrium
between gas pressure and rest--mass self--gravity, to a very general
equilibrium between energy--momentum flux and total self--gravity.

% \bibliography{xxx}
% \bibliographystyle{apsrev}

\end{document}